\documentclass[letterpaper]{article} %
\usepackage{aaai2026}  %
\usepackage{times}  %
\usepackage{helvet}  %
\usepackage{courier}  %
\usepackage[hyphens]{url}  %
\usepackage{graphicx} %
\usepackage{natbib}  %
\usepackage{caption} %
\usepackage{makecell} %
\usepackage{tikz}
\usepackage{algorithm, algpseudocode}
\usepackage{amsmath}
\usepackage{newfloat}
\usepackage{listings}
\DeclareCaptionStyle{ruled}{labelfont=normalfont,labelsep=colon,strut=off} %
\floatstyle{ruled}
\newfloat{listing}{tb}{lst}{}
\floatname{listing}{Listing}
\usepackage{bibentry}

\title{\emph{SubstratumGraphEnv}: Reinforcement Learning Environment (RLE) for Modeling System Attack Paths}

\author {
    Bahirah Adewunmi\textsuperscript{\rm 1},
    Edward Raff, PhD\textsuperscript{\rm 2},
    Sanjay Purushotham, PhD\textsuperscript{\rm 1}
}
\affiliations {
    \textsuperscript{\rm 1} University of Maryland, Baltimore County\\
    \textsuperscript{\rm 2} CrowdStrike\\
    b280@umbc.edu, edward.raff@crowdstrike.com, psanjay@umbc.edu
}

\begin{document}
\maketitle
\begin{abstract}
Automating network security analysis, particularly the identification of potential attack paths, presents significant challenges. Due in part to the sequential, interconnected, and evolutionary nature of system events which most artificial intelligence (AI) techniques struggle to model effectively. This paper proposes a Reinforcement Learning (RL) environment generation framework that simulates the sequence of processes executed on a Windows operating system, enabling dynamic modeling of malicious processes on a system. This methodology models operating system state and transitions using a graph representation. This graph is derived from open-source System Monitor (Sysmon) logs. To address the variety in system event types, fields, and log formats, a mechanism was developed to capture and model parent-child processes from Sysmon logs. A Gymnasium environment (\emph{SubstratumGraphEnv}) was constructed to establish the perceptible basis for an RL environment, and a customized PyTorch interface was also built (\emph{SubstratumBridge}) to translate Gymnasium graphs into Deep Reinforcement Learning (DRL) observations and discrete actions. Graph Convolutional Networks (GCNs) concretize the graph’s local and global state, which feed the distinct policy and critic heads of an Advantage Actor-Critic (A2C) model. This work's central contribution lies in the design of a novel deep graphical RL environment that automates translation of sequential user and system events, furnishing crucial context for cybersecurity analysis. This work provides a foundation for future research into shaping training parameters and advanced reward shaping, while also offering insight into which system events attributes are critical to training autonomous RL agents.
\end{abstract}

\section{Introduction}
The overwhelming majority of cyber incidents stem from human factors, with 88\% of breaches tied to human error (Hancock 2022) and 80\% involving compromised credentials (Hylender et. al 2025). A significant portion of these attacks exploit misconfigurations, as evidenced by 23\% of security incidents being traced back to cloud misconfigurations alone (Sentinel One 2025). Automating network security analysis, particularly mapping potential attack paths, is also challenging due to the complex, sequential nature of system events that are complicated by the interconnected and often exceptionally long sequences of system events involved. Traditional AI modeling also struggles, not simply because the data is sequential, but because the critical attack sequences' attributes are highly dimensional and more intricately linked than what typical sequence-based AI models are designed to handle effectively. While conventional sequence models like Recurrent Neural Networks (RNNs) and their variants (e.g., Long Short-Term Memory networks, Gated Recurrent Units) can process temporal data, they often fall short in capturing the complex, non-linear, and often graph-structured dependencies inherent in system process logs. 

More sophisticated approaches are needed, such as Graph Neural Networks (GNNs), which inherently model relational data, Transformers with their powerful attention mechanisms for capturing long-range dependencies and rich contextual embeddings, or Deep Reinforcement Learning (DRL) which can learn optimal policies by interacting with and exploring these complex, dynamic environments.

While various applications of RL and graph-based methods exist in cybersecurity, our work distinctively addresses the challenge of directly modeling raw, high-dimensional operating system event data from Sysmon logs as a dynamic, tensor-based graph for DRL. Our primary contribution is the development of a novel deep graphical based RL environment that converts raw, high-dimensional operating system event data from Sysmon logs into a tensor-based graph suitable for training neural network-based autonomous agents. This enables the training of RL agents to understand and learn concrete system process paths, a critical functionality for objectively auditing and evaluating system security posture.

The insights gained from this study will be instrumental in designing more complex observation states and designing and training RL algorithms to understand the subtle, yet critical, temporal changes attributable to malicious and benign actions for future, more advanced analysis and detection efforts.

\section{Literature Review}
While various applications of RL and graph-based methods exist in cybersecurity, our work distinctively addresses the challenge of directly modeling high-dimensional process level event data. Our literature review revealed a significant gap: there are no directly comparable methods that either identify attack paths at the application or process level, abstract attack paths from observed process dependencies, or generate data structures enabling autonomous agents to learn and traverse system process paths. Due to the observed scarcity of enterprise host event logs, a constraint often imposed by privacy and security concerns, all related studies are forced to RL action spaces based on network flows or human-curated expertise (e.g., Intrusion Detection System [IDS] rules and known vulnerabilities).

Existing methodologies for mapping low-level system events to RL states, actions, and rewards vary widely and often lack a standardized approach. Many approaches abstract away the underlying system processes. For instance, Maeda and Mimura (2020) model post-exploitation tactics using a composite of ten system parameters rather than granular processes, while Harries et al. (2020) employ symbolic trees to model user interface interactions. These studies operate on a different modality of system activity, distinct from the raw event sequences we model.

In the context of attack detection, several works apply RL but at a different level of abstraction. Deokar and Hazarnis (2012) use RL to correlate log files from multiple sources for attack confirmation, and Xu and Xie (2005) use RL for intrusion detection by modeling system call sequences as a Markov reward process. While valuable, these methods do not involve an agent taking explicit exploratory actions in a dynamic, graph-based environment. Similarly, in penetration testing, Yang et al. (2022) and Elderman et al. (2017) developed RL frameworks to discover attack paths, but their state representations are typically matrices of network and host information, not the detailed, graph-structured relationships of processes and events derived from raw system logs.

More recently, research has combined DRL with graph representations for cybersecurity tasks, but these generally operate at higher levels of abstraction. Surveys by Gueriani et al. (2024) and Jamshidi et al. (2025) on DRL-based Intrusion Detection Systems (IDS) in IoT highlight the use of DRL and sometimes GNNs, but they often rely on network flow features or aggregated data rather than granular, raw operating system (OS) events. Similarly, works like Cody et al. (2022) and Li et al. (2024) leverage RL to identify attack paths, but their graph representations model network topologies and vulnerabilities, not the dynamic process relationships that are central to our study. While Li et al. (2024) GCNs, it focuses on data integrity in specific cyber-physical systems and does not address the conversion of general OS event streams into a tensor-based graph for learning system behaviors. Wang et al. (2025) also advance RL for network defense, but they do not delve into the granular modeling of host-based system processes from raw logs.

A critical challenge across these RL applications is designing an effective reward function, as highlighted by Zheng et al. (2021). Training agents in complex, sequential environments faces issues of sparse and delayed rewards, which is exacerbated in large state spaces. Furthermore, while  Kujanpää et al (2021) investigate privilege escalation, translating these high-level security behaviors to the low-level events in logs like Sysmon remains complex. The development of simulation environments is essential for applying RL to security, but general graph-based training frameworks typically do not capture the specific semantics and sequential dependencies inherent in realistic OS event data (Ergun et al., 2023; Zhao et al., 2020; Liu et al., 2025).

\section{Method}
\subsection{Datasets}
Reliably configured, high-fidelity enterprise host event logs remain exceptionally scarce. Kenyon et al. (2020) found this is compounded by issues like an inconsistent central registry, rapid data obsoletion due to evolving threats, and heavy anonymization in older datasets. Scarce access to parent-child OS event datasets is exacerbated by not being a default logging configuration for Windows and Linux systems.

As indicative of these data constraints, this study leverages two open-sourced Sysmon data sources generated from cyber attack simulations. For Windows OS, Sysmon excels in capturing detailed parent-child process relationships. The first source is named BRAWL, described by Kemmerer and Wampler (2018) as a cloud-hosted enterprise network employing MITRE's Caldera as an offensive actor. The events from BRAWL map directly to the MITRE Cyber Analytics Repository (CAR), enabling the capture of various post-compromise adversary behaviors like credential dumping and lateral movement. To diversify the evaluation of our methodology's robustness, specifically in constructing the network and Gymnasium graph, the Cerberus Traces dataset (Pratomo, 2023) was also incorporated. This dataset provides additional Sysmon sequential events, collected from real Windows environments, encompassing both malicious and benign software executions.

\subsection{Graph Design}
Our approach models system activity as a dynamic directed graph derived from processed Sysmon logs. The fundamental purpose of this design is twofold: first, to accurately and structurally capture the causal process dependencies (e.g., parent-child relationships) that are often obscured or missing in native OS logs, thereby revealing complex attack paths enabled by system misconfigurations or overlooked security parameters. Second, the graph structure provides a formally defined state representation that aligns with the Markov Decision Process (MDP) framework. This transformation converts raw log data into a structured and traversable format, which is essential for creating a valid deep Reinforcement Learning (DRL) environment and reward function where an autonomous agent can learn to identify and traverse the sequences that constitute an adversarial event. Analysis of the generated graphs revealed that the subsequent reward function must prioritize and incentivize long-range traversal and the discovery of critical, sequential dependencies across a dispersed set of nodes.

\begin{enumerate}
  \item {\bf Nodes}: Represent discrete system states or entities (e.g., running processes identified by executable path and identifier (ID), executed commands).
  \item {\bf Edges}: Signify transitions between states, representing a causal link like a parent process creating a child process or a process executing a command.
  \item {\bf Attributes}: Edges include attributes such as interaction frequency (used as weight) and contextual metadata. Nodes hold attributes like process name, parent process ID, integrity level, event duration, process ID, image path, host name, file path, etc.
\end{enumerate}

The parent-child relationships were established using Sysmon Event ID 1 (Process Creation) as the parent process. Child nodes are derived from Sysmon Event IDs such as Event ID 2 (File creation time changed), 5 (Process terminated), 11 (File created), 12 (Registry object created/deleted), and 13 (Registry value set).  The BRAWL and Cerberus Traces datasets were processed to extract these events as nodes.

\begin{figure}[t]
\includegraphics[width=.9\columnwidth]{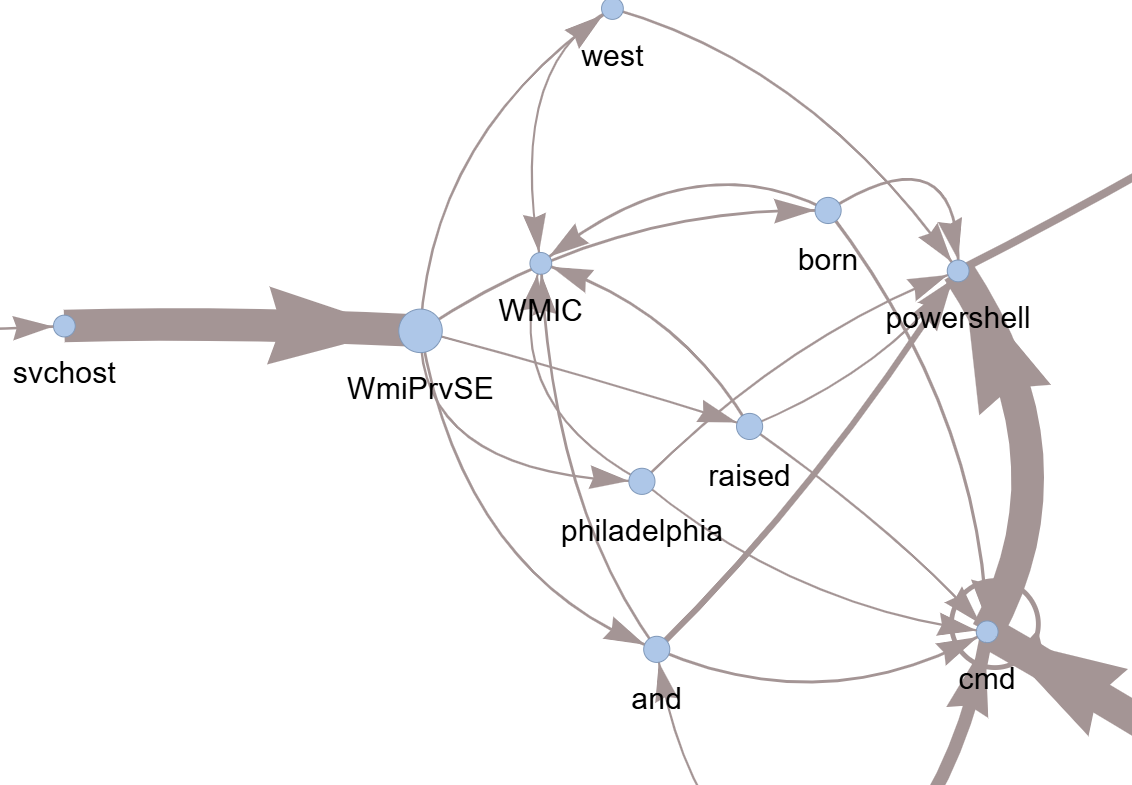}
\caption{This partial network snapshot of the BRAWL dataset reveals the sparse and disassortative nature of the system's process graph. The node labels—ranging from standard Windows processes like \emph{svchost} to strings from the "Fresh Prince" theme—are raw identifiers extracted directly from the dataset's Sysmon logs. While frequent processes like \emph{powershell} have numerous connections (thick edges), the low clustering demonstrates that an RL agent cannot rely on dense local patterns, but must instead learn to navigate these non-intuitive, sparse pathways to identify the full attack chain.}
\captionsetup{justification=centering}
\label{fig:brawl}
\end{figure}

 Figure \ref{fig:brawl} illustrates the directional parent and child relationships in the Sysmon logs, where the source node is the parent process and the terminal node is the child process.  Visually observed low clustering suggests that a reward function relying solely on dense local event patterns would be ineffective. Instead, the RL state representation should focus on individual events and their immediate sequential relationships, and the agent must learn to traverse longer, sparser pathways to achieve security-relevant goals.

The collected and filtered event data is transformed into NetworkX directional graphs, represented as \emph{DiGraph} objects. This process interprets the relationships identified in logs as directed edges between nodes representing system entities. Node and edge attributes are extracted from log fields and attached to the corresponding graph elements. These NetworkX graphs provide a structured representation of system event sequences.

Developed as a custom Gymnasium environment, \emph{SubstratumGraphEnv} preserves the graph's connectivity information and models the state space as a \emph{DiGraph}. It facilitates the interface between graph representations and DRL frameworks by converting the internal NetworkX graph state into the \emph{gymnasium.spaces.Graph} dict, giving the node and edge attributes a standardized dictionary layout.Each entry stores one component of the graph snapshot: node feature matrix, edge index pairs, plus scalar counts of the number of nodes and edges.

\begin{figure}[h]
\centering
\includegraphics[width=.9\columnwidth]{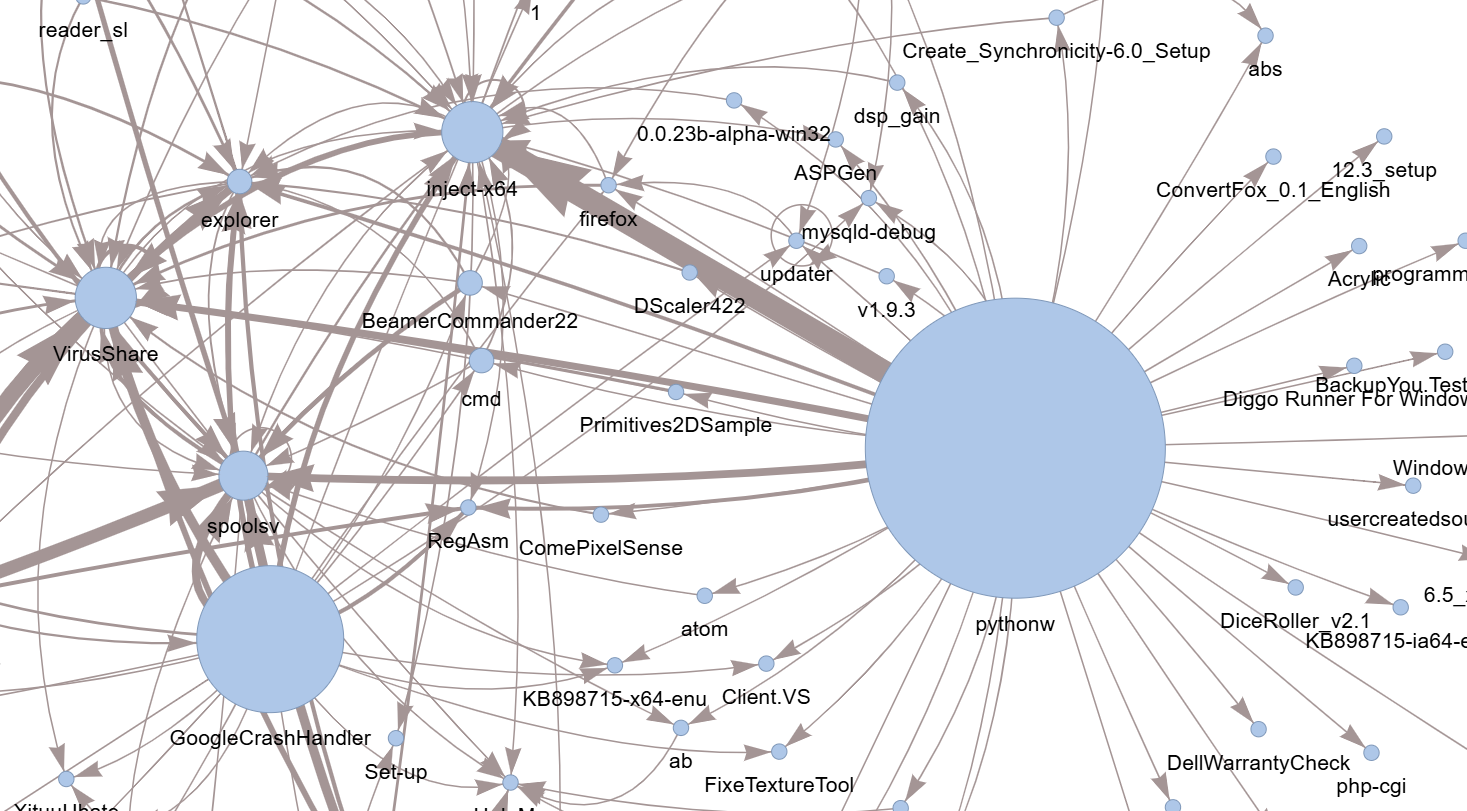}
\caption{This snapshot of Cerberus Traces reveals a dense, complex process graph with a few processes, like \emph{pythonw}, acting as central hubs for activity. Despite this high connectivity, the network's low clustering shows that critical attack sequences are not confined to local, dense event patterns. Therefore, the RL environment must guide the agent through a large, distributed state space to discover significant sequential connections.}
\label{fig:cerberus}
\end{figure}
\subsection{Graph Analysis}
We conducted a structural analysis of the Sysmon event graphs on the BRAWL and Cerberus Traces datasets. We observed high sparsity, low clustering coefficients, and disassortative mixing patterns (as detailed in Tables \ref{tab:Table 1} and \ref{tab:Table 2}), revealing that attack sequences exploiting process dependencies do not form dense, local clusters. These network characteristics reflect the attack chain process, which is that a cyber environment is distinguished by sparse and delayed reward challenges. In terms of autonomous agent behavior, this often requires an attacker to execute numerous, sequential actions over an extended duration to achieve a goal. Consequently, this analysis calls for a nuanced reward function capable of guiding an agent to learn and traverse long, distributed, and critical paths within a vast state space.

After dropping childless parent nodes, the graph derived from the BRAWL dataset contained 104 nodes and 227 edges, exhibiting a density of $2.12\times{10}^{-2}$. The Cerberus Traces dataset yielded a larger graph with 6,033 nodes and 7,446 edges, with a density of $2.05\times{10}^{-4}$. Analysis of each graph's Clustering Coefficient, which measures local interconnectedness (the likelihood of neighbors of a node also being connected), revealed predominantly low values across both networks and highlighted limited dense and local clusters of events. The BRAWL network generally showed low average nearest neighbor degrees, particularly for high-degree nodes, suggesting a disassortative mixing pattern where frequently occurring events connect to less frequent ones. 
\begin{table}[h]
    \begin{tabular}{c c | c c}
        \multicolumn{2}{c|}{\textbf{Cerberus Traces}} & \multicolumn{2}{c}{\textbf{BRAWL}} \\
        \hline
        \thead{Key \\Degree (k)} & \thead{Average \\Nearest\\Neighbor \\Degree (${k}_{nn}$)} & \thead{Key \\Degree (k)} & \thead{Average \\Nearest\\Neighbor \\Degree (${k}_{nn}$)} \\
        \hline
        1    & 487.94 & 1    & 0.10 \\
        2    & 103.11 & 2    & 0.77 \\
        3    & 152.66 & 3    & 0.46 \\
        4    & 55.98  & 4    & 0.67 \\
        15   & 63.08  & 5    & 0.34 \\
        30   & 0.36   & 9    & 0.45 \\
        55   & 2.73   & 15   & 0.18 \\
        97   & 2.40   & 18   & 0.06 \\
        109  & 2.69   & 20   & 0.30 \\
        152  & 13.22  & 24   & 0.01 \\
        300  & 0.83   & 56   & 4.35 \\
        23   & 4.55   & 62   & 0.01 \\
        24   & 7.41   & ~23-24 & ~0.01 (k=24) \\
        Max (2566) & 0.58 & \multicolumn{2}{c}{\textemdash}\\
        \hline
    \end{tabular}
    \caption{ The highly varied, often low Average Nearest Neighbor Degree and the disassortative patterns where high-degree nodes connect to low-degree neighbors, demonstrate that rewards cannot simply rely on local, dense clusters of events}
    \label{tab:Table 1}
\end{table}
The Cerberus Traces network presented a more complex pattern. While some high-degree nodes were disassortative, many lower-degree nodes connected to very high-degree neighbors. This indicates that certain peripheral events might strongly link to central system activities. These connectivity patterns highlight the varied structural contexts an RL agent must learn to navigate. Comparison of key degree and Average Nearest Neighbor Degree between nodes for Cerberus Traces and BRAWL Datasets in Table \ref{tab:Table 1}, demonstrate that the reward function must be meticulously designed to propagate signals across nodes with vastly different connectivity. The policy will need a reward to effectively guide the agent to traverse long, sparse, and non-intuitive pathways to achieve security-relevant goals.

In-degree Centrality analysis identified nodes that receive a high number of incoming connections (representing reliance or targets of interaction). Table \ref{tab:Table 2} lists the top nodes by in-degree centrality for both datasets. The BRAWL network featured core system utilities like \emph{conhost}, \emph{cmd}, and \emph{powershell} with relatively high in-degree centrality, suggesting a more centralized interaction pattern around these processes. 

In contrast, the top nodes in the larger Cerberus Traces network exhibited significantly lower in-degree centrality scores, suggesting a more distributed interaction pattern and a lack of highly-centralized, common entry points. This has a direct impact on RL environment design, as a sparse, low-centrality graph structure complicates the design of a simple reward mechanism. A reward function based solely on reaching a single, highly-connected node would be ineffective. Instead, an effective reward structure must be capable of rewarding the agent for navigating a larger, more dispersed set of critical nodes, or for achieving a cumulative effect across multiple less connected nodes, reflecting the more distributed nature of the interactions within the Cerberus dataset. 

\begin{table}
  \label{Table 2}
  \centering
\resizebox{\columnwidth}{!}{%
  \begin{tabular}{lcc}
    \multicolumn{1}{c}{\textbf{Process/Node Name}} & \multicolumn{1}{c}{\textbf{In-Degree Centrality Value}} & \multicolumn{1}{c}{\textbf{Dataset}}\\
    \hline
    conhost      & 0.13600    & BRAWL  \\
    Cmd          & 0.06800    & BRAWL  \\
    Powershell   & 0.05800    & BRAWL  \\
    Reg          & 0.05800    & BRAWL  \\
    Net          & 0.04900    & BRAWL  \\
    WMIC         & 0.04900    & BRAWL  \\
    GoogleUpdate & 0.02900    & BRAWL  \\
    ProxyEnable  & 0.00862    & Cerberus Traces  \\
    inject-x64   & 0.00381    & Cerberus Traces  \\
    is32bit      & 0.00381    & Cerberus Traces  \\
    \hline
  \end{tabular}%
}
  \caption{The significantly higher centrality scores in BRAWL (e.g., \emph{conhost}, \emph{cmd}) versus Cerberus Traces reveal two distinct interaction patterns: centralized in BRAWL and distributed in Cerberus. This structural difference confirms the need for a reward function that can target dispersed critical nodes, not just single, highly connected targets.}
\label{tab:Table 2}
\end{table}

Collectively, these graph properties (sparsity, varied degree connectivity patterns, low clustering) informed the design of the RL environment, highlighting the necessity for: 1. A state representation capable of handling sparse, potentially high-dimensional input; 2. A reward function that can capture distributed patterns and guide learning over sequential, non-locally clustered events; and 3. RL algorithms suited for exploration in environments with limited local connectivity and large state/action spaces.

\subsection{Reinforcement Learning Environment Design}
To circumvent Gymnasium library's incompatibility with \emph{torchrl.envs.GymEnv}, we developed \emph{SubstratumBridge}. When this paper was written, PyTorch's \emph{gymenv} wrapper is incompatible with graphs built on Gymnasium version 1.0 or newer due to unpredictable behavior. /emph{SubstratumBridge} converts graph attributes into a PyTorch Tensor format, enabling neural network training on system process attributes like process event integrity level, duration of events, and user information. This focus on granular system event data, prioritizes learning security objectives aligned to attack detection and attack path characteristics, while also accounting for the sequential dependencies within the graph. This improves upon more abstract methods seen in prior work, which often abstract away underlying system processes or focus on user interface interactions.

The Markov Decision Process (MDP) offers a powerful framework for modeling network attack paths, treating running processes as discrete points in time, and the transitions between these states as being driven by logically linked, child events. Parent-child event relationships are prime examples of such causal links, encapsulating critical information for predicting and understanding the flow of events within a system. 

MDP is is typically defined by a tuple ($S$,$A$,$P$,$R$,$\gamma$):
\begin{equation}
V^*(s) = \max_{a \in \mathcal{A}} \sum_{s' \in \mathcal{S}} \mathcal{P}(s'|s, a) \left[ \mathcal{R}(s, a, s') + \gamma V^*(s') \right]
\end{equation}

In our experimental setup, the state ($s\epsilon S$) at any given time is defined by the current node (process) in the Sysmon event graph. The action ($a\epsilon A$) is a discrete selection of target nodes, with validity checks against a node adjacency matrix. $V^*(s')$ is the maximum expected cumulative future reward that can be obtained starting from state $s'$.

The reward function ($R(s,a,{s}')$) was designed to guide the agent through sparse, complex graph environments. The baseline reward algorithm provided dense learning signals via per-step penalties, severe penalties for invalid, non-adjacent moves, and graduated rewards for integrity level progression and in-degree centrality.

\begin{algorithm}
\caption{Baseline Reward Function $\mathcal{R}(s, a, s')$}
\label{alg:reward_function}
\begin{algorithmic}[1]
\State \textbf{Input:} Current State $s$, Action $a$ (Target Node $s'$), Graph $\mathcal{G}$, Constants $\alpha, \beta, R_{step}, R_{term}, R_{escalate}, R_{downgrade}$
\State \textbf{Output:} Reward $R$
\State $R \leftarrow R_{step}$ \Comment{Constant negative step penalty for efficiency}
\If{State $s$ is Invalid \textbf{or} Target $s'$ is Out of Bounds}
    \State $R \leftarrow -1.0$ 
    \State \textbf{Return} $R$
\EndIf

\If{Edge $(s, s')$ is NOT present in $\mathcal{G}$}
    \State $R \leftarrow R_{step} - 0.5$
    \State \textbf{Return} $R$
\Else
    \State $w \leftarrow \text{Edge Weight}(s, s')$
    \State $R \leftarrow R + w$ 
    \State $R \leftarrow R - (\alpha \cdot w)$ 
    \State $I_s \leftarrow \text{Integrity Level}(s)$
    \State $I_{s'} \leftarrow \text{Integrity Level}(s')$
    \If{$I_{s'} > I_s$}
        \State $R \leftarrow R + R_{escalate}$
    \ElsIf{$I_{s'} < I_s$} 
        \State $R \leftarrow R + R_{downgrade}$
    \EndIf
    \State $C_{s'} \leftarrow \text{In-Degree Centrality}(s')$
    \State $R \leftarrow R + (\beta \cdot C_{s'})$ 
    \If{$s'$ is the Terminal Node $s_{term}$}
        \State $R \leftarrow R + R_{term}$ 
        \State $\text{Episode Terminated} \leftarrow \text{True}$
    \EndIf
\EndIf
\State \textbf{Return} $R$
\end{algorithmic}
\end{algorithm}

Algorithm \ref{alg:reward_function} outlines the reward function. This function was designed to directly address the sparsity, low clustering, and varied degree connectivity observed in the Cerberus Traces and BRAWL graph analysis. A constant negative step penalty incentivizes efficiency across the long, distributed pathways; concurrently, strong penalties of $-1.0$ and $-0.5$ deter incorrect actions and enforce the strict adjacency rules of the graph. To overcome the prevalence of benign noise and earn attributes important for detecting attack paths, edge traversal penalties discourage frequently selecting the same edges, while integrity level bonuses ($+100$ escalation) and target node centrality bonuses specifically guide the agent toward significant network processes. These attributes were added to the reward algorithm to compensate for the distributed interaction patterns that complicate learning in sparse graphs. Finally, a large $+5000$ bonus reinforces reaching the terminal node set at the start of the experiment, providing a non-local objective and setting a reward signal to combat environment sparsity.
\usetikzlibrary{shapes.geometric, arrows.meta, positioning}

\begin{figure}
  \centering
\begin{tikzpicture}[
  every node/.style={font=\small, align=center},
  startend/.style={draw, thick, ellipse, fill=blue!6,
                   text width=3.2cm, minimum height=0.9cm},
  process/.style={draw, thick, rectangle, fill=blue!4,
                  text width=3.4cm, minimum height=0.95cm},
  decision/.style={draw, thick, diamond, aspect=2, fill=orange!15,
                   text width=3.4cm, minimum height=1.1cm},
  envblock/.style={process, fill=green!12},
  bridgeblock/.style={process, fill=green!25},
  arrow/.style={-Latex, thick}
]

\node[startend] (start) {Sysmon relations ingested (CSV / JSON)};
\node[process, below=9mm of start] (normalize) {Normalize relations};
\node[process, below=9mm of normalize] (graph) {Construct NetworkX DiGraph};
\node[process, below=9mm of graph] (encode) {Encode + pad attributes};
\node[envblock, below=9mm of encode] (env) {Run SubstratumGraphEnv (Gym)};
\node[bridgeblock, below=9mm of env] (bridge) {Wrap with SubstratumBridge (TorchRL)};
\node[startend, below=9mm of bridge] (end) {TorchRL-ready graph experience};

\draw[arrow] (start) -- node[right, font=\scriptsize]{structured relations} (normalize);
\draw[arrow] (normalize) -- node[right, font=\scriptsize]{clean dict-of-dicts} (graph);
\draw[arrow] (graph) -- node[right, font=\scriptsize]{attributed graph} (encode);
\draw[arrow] (encode) -- node[right, font=\scriptsize]{fixed-length feature vectors} (env);
\draw[arrow] (env) -- node[right, font=\scriptsize]{Gym obs + rewards} (bridge);
\draw[arrow] (bridge) -- node[right, font=\scriptsize]{TensorSpec + TensorDict outputs} (end);

\end{tikzpicture}
  \caption{High-level pipeline from raw Sysmon relations to TorchRL-ready graph observations.}
\end{figure}
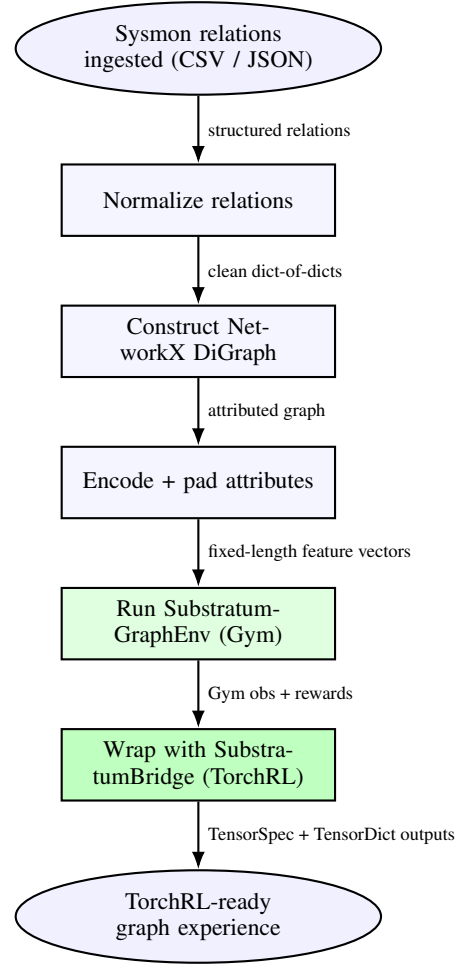

Subsequent refinement increased penalties for invalid moves by a factor of 10 to strongly discourage undesirable actions. Concurrently, the penalty for moving to a lower integrity level was substantially reduced by a factor of 100 to offer the agent more flexibility. Most notably, the edge weight frequency penalty was replaced with a penalty based on the frequency of selecting the same node, which progressively increases each time a node is revisited within an episode. 

\subsection{Graph-Based Actor-Critic Model Design}
A graph-based Advantage Actor-Critic (A2C) architecture design was chosen to address the need to process the sparse, non-local dependencies inherent in the graph networks. GCN layers are critical, as they capture both local features and global graph topology by processing the features of each node and iteratively aggregating information across the graph's entire node adjacency matrix. The output from the GCN layers is consolidated into a single, fixed-size vector representing the overall graph state using global pooling. This process, which transforms the dynamic graph state into the tensor format, validates the feasibility of the designed DRL environment and observation space for objective, process-level security analysis using a basic RL agent.

\begin{enumerate}
  \item {\bf Initial Node Embedding}: Represent discrete system states or entities (e.g., running processes identified by executable path and identifier (ID), executed commands).
  \item {\bf Graph Convolutional Layers}: Two PyTorch GCNConv layers, both size $m$. Process the node embeddings by aggregating information from neighbors to capture local and wider graph topological context. Rectified Linear Unit (ReLU) activation functions are applied after these layers, and dropout is included for regularization.
  \item {\bf Graph Consolidation}: A mean global pooling layer aggregates the learned node representations from the GCNs into a $m$ sized vector representing the entire graph state.
  \item {\bf Actor}: Linear layer. Maps the consolidated graph representation to a vector of unnormalized scores or logits for selecting each possible discrete action space ($n$ possible actions), defining the agent's policy.
  \item {\bf Critic}: Linear layer. Maps the consolidated $m$ sized graph representation to a scalar value, the model's estimate of the expected cumulative future reward obtainable from the current state.
\end{enumerate}

The GCN's training process utilizes the Adam optimizer to minimize a combined loss function. Adam is a popular and robust choice for training many types of deep neural networks, because it combines the benefits of root mean square propagation (RMSprop) and momentum. The loss function has three main components:
\begin{enumerate}
  \item {\bf Policy Loss ($L_p$)}: Mnih et al (2016) provide the algorithm for policy loss, which is formulated to increase the probability of actions that result in a positive advantage and decrease the probability of actions that result in a negative advantage. For a batch of N samples (state-action pairs), the policy loss is typically calculated as:
  \begin{equation}
  L_{p}=-\frac{1}{N}\sum_{i=1}^{N}{\log\pi(a_i|s_i)A(s_i,a_i)}
  \end{equation}
  \item[] Where $\pi(ai|s_i)$ the probability of taking action $a_i$ in state $s_i$ according to the actor's policy, and $A(s_i,a_i)$ the estimated advantage of taking action $a_i$ in state $s_i$.
  \item {\bf Value Loss ($L_v$)}: The value loss uses PyTorch's (Barnes 2021) implementation of Mean Squared Error (MSE) loss. It measures the difference between the critic's estimate and the actual observed returns from the environment. The strength of the value loss is regulated by a value loss coefficient. For a batch of N states, the value loss ($L_v$) is:
  \begin{equation}L_v=\frac{1}{N}\sum_{i=1}^{N}(V(s_i)-R_i)^2\end{equation}
  \item[] Where $V(s_i)$ is the value estimated by the critic for state $s_i$, and $R_i$ is the calculated target return (discounted cumulative reward) from state $s_i$.
  \item {\bf Entropy Regularization ($L_e$)}: PyTorch's entropy term is included to encourage exploration by promoting diversity in the agent's action choices (Fritz 2017). This helps prevent the policy from converging too quickly, becoming too deterministic, or forming a suboptimal. The strength of regularization is controlled by a coefficient.
  \begin{equation}L_{e}=-\sum_{a\in A}{\pi(a|s_i)\log(\pi(a|s_i))}\end{equation}
  \item[] Where $\pi(a|s_i)$ is the probability of taking action $a$ in state $s_i$ according to the policy $\pi$, A is the set of all possible actions, and $\beta$ is the entropy coefficient, controlling the strength of the regularization.
\end{enumerate}
The total loss minimized by the optimizer is a weighted sum of these components:
\begin{equation}L_{Total}=L_p+c_{v}L_v+c_{e}L_e\end{equation}

\subsection{Experiment Design}
The fundamental purpose of this experimental design is to test and measure the feasibility and training stability of the DRL environment and its observation space.  To select a plausible attack path to a process of interest, the longest path in each network were programmatically identified. The selection process utilized NetworkX's \emph{dag-longest-path} function on each network's condensation graph due to directional cycles within both graphs. For the BRAWL dataset, this defined a path from the \emph{TrustedInstaller} process to a IPv6 address. In the Cerberus Traces dataset, a path was identified that originated from the \emph{execsc} process and terminated at \emph{Procmon}.

Experiment runs were executed on a Debian 12 Operating System with two terabyte (TB) disk capacity, 64 gigabytes (GB) of Randomized Access Memory (RAM), an NVIDIA Geoforce RTX 3090 GPU card with 24 GB virtual RAM, and an AMD Ryzen 9 7950X Computing Processing Unit (CPU) with 16 Cores,and 32 logical processors. 

Experiments aimed to assess the RL agent training stability within the graphical environment and explore the impact of various model hyper-parameters on the RL environment's performance. This resulted in experiment runs over predefined sets of values for key hyper-parameters. Specifically, the experiments are initialized for each unique combination of these hyperparameter values:
\begin{enumerate}
    \item\emph{model-output-dim}: The dimensionality of the GCN's output.
    \item\emph{value-loss-coef}($c_v$): Controls the weighting of the value loss component in the overall loss function, with options of 0.5, 1.0, and 2.0.
    \item\emph{entropy-coef} ($c_e$): Influences the exploration-exploitation trade-off by weighting the entropy regularization term, using values 0.001, 0.01, 0.05, or 0.1.
    \item\emph{learning-rate}: Determines the step size at which the model's weights are updated during training, exploring rates 0.0001, 0.001, 0.005, and 0.01.
    \item\emph{total-training-steps}: Defines the total number of training steps for each run (1000, 2000, or 5000).
    \item\emph{node-frequency-penalty-scale}: This parameter, varied at 0.5, 5.0, and 20.0, adjusts the penalty applied for revisiting nodes within an episode, promoting exploration of the network.
    \item\emph{num-parallel-envs-options}: The number of parallel environments leveraged in experiment runs varied between 2, 3, 12, 16, and 32
\end{enumerate}

Other hyperparameters such as the discount factor ($\gamma = 0.99$), and the number of steps before each update (\emph{trainer-n-steps = 3}) are kept constant across all experiments. We conducted a comprehensive hyperparameter sweep, running multiple combinations on the BRAWL dataset. In contrast, due to the significantly larger size and longer runtime required by the Cerberus Traces dataset, we performed one combination of hyperparameters. 

\section{Findings}
The primary objective of this work is to develop and validate a stable DRL environment capable of modeling and analyzing complex Sysmon logs as a graph. One report collected the minimum and maximum total episode awards from the runs, demonstrating the environment's stability and integrity during training. Another report tracks and aggregates training metrics: computing the global step count, episode count, average run reward, average run policy loss, average run value loss, average run entropy, and reward and steps for the current run.

The run metrics are averaged by aggregating data from all parallel workers. For example, the average policy loss is computed by summing the total policy loss and dividing by the total steps from all workers for that run. The policy loss measures how well the agent's policy is learning to take actions that maximize rewards. The value loss measures the discrepancy between the critic's estimated value of a state and the actual observed returns from the environment

\subsection{Cerberus Traces Environment}
\begin{figure}
    \centering
    \includegraphics[width=0.99\linewidth]{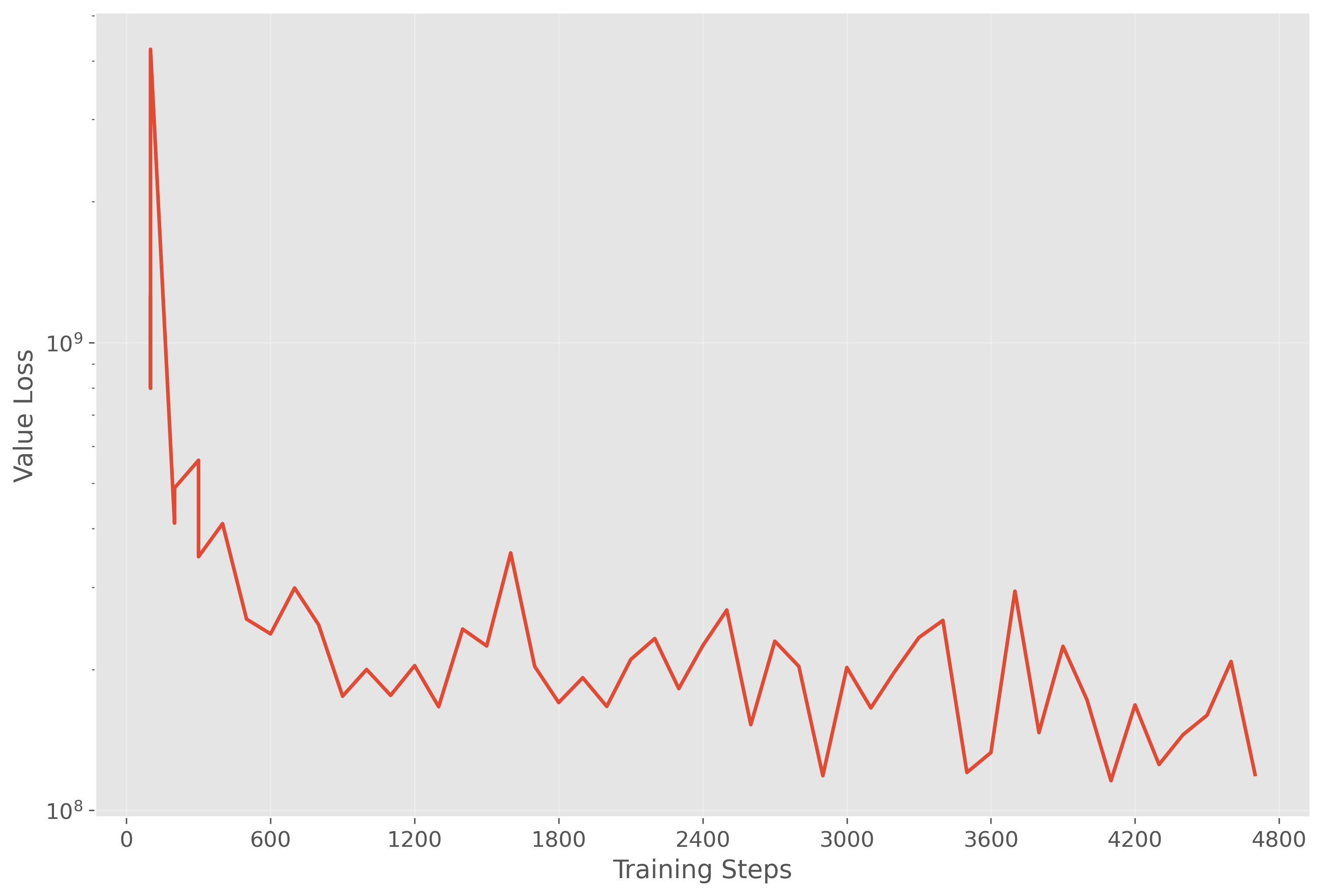}
    \caption{Value loss versus training steps on the Cerberus DRL environment. Decreasing value loss indicates that critic is learning to predict value accurately}
    \label{fig:Value Loss Cerberus}
\end{figure}

With the Cerberus Traces dataset, we observed a successful experiment run that demonstrated the stability and functionality of the DRL environment. As seen in Figure \ref{fig:Value Loss Cerberus}, the value loss for the Cerberus DRL environment exhibits a rapid initial decay, dropping steeply from a high starting point before stabilizing into a fluctuating, yet low-magnitude, pattern after approximately 1200 training steps. This collectively indicates that the Critic network successfully learned to predict the value of the highly sparse and complex graph states with reasonable accuracy. Our DRL environment completed 4,710 training steps over a duration of 10 hours and 42 minutes, demonstrating the environment's long-term stability and its capacity to support extended training and the iterative training process of an RL agent. 

Once parallelization was introduced, it unsurprisingly created a memory bottleneck influenced by the number of model output dimensions and the number of parallel environment workers. The Cerberus Traces dataset was able to complete 24 to 120 steps per run with two workers and a model output dimension of 64, averaging 537 seconds per run without parallelization. But holding the model's hyper-parameters the same, the number of steps completed dropped, varying from 18, 72, or 81 steps per run, with an average duration of 292 seconds. All parallel runs with the Cerberus Traces network prematurely terminated due to out-of-memory errors on the GPU.

\subsection{BRAWL Environment}
Table \ref{tab:comparison_metrics} illustrates that the Cerberus Traces DRL environment, characterized by more distributed node interactions, yielded stable and rapidly converging results (zero loss standard deviation). Conversely, the BRAWL environment, despite being smaller, demonstrated extreme training instability (high average loss and high standard deviation, e.g., $1.01\times 10^{13}$). The BRAWL DRL environment confirmed that its relatively more sparse structure and low node degree present a fundamentally harder problem for the RL agent. Regardless, across various hyper-parameter combinations and training epochs, the BRAWL environment consistently produced valid observations and managed episode state transitions. 

\begin{table}[h]
    \centering
    \begin{tabular}{lcc}
        \hline
        & \textbf{BRAWL} & \textbf{Cerberus} \\
        \hline
        Avg Reward (mean) & -18.42 & -4.66 \\
        Avg Segment (std) & 63.83 & 0.76 \\
        Avg Policy Loss (mean) & 0.09 & 0 \\
        Avg Policy Loss (std) & 26.42 & 0 \\
        Avg Value Loss (mean) & 1.21E+11 & 4.58E+08 \\
        Avg Value Loss (std) & 1.01E+13 & 6.87E+08 \\
        \hline
    \end{tabular}
    \caption{Averaged Parallelized Worker Training Metrics for the DRL Environments. The Cerberus environment demonstrates feasibility and stability with negligible loss variance. In contrast, the high loss standard deviations in BRAWL ($1.01\times 10^{13}$) confirm that its sparse graph structure presents a challenging stability problem for the DRL agent.}
    \label{tab:comparison_metrics}
\end{table}

542 experiments with the BRAWL DRL environment were completed sequentially. Holding the number of parallel workers constant, the runtime lengths for the experiments varied from 1 to 3  seconds, demonstrating the environment's reliable and predictable performance under various hyper-parameter configurations.

\section{Conclusion}
This paper introduces a new approach to automated security analysis and attack path mapping using DRL on a graph of system events derived from Sysmon logs. Moving beyond the limitations of current security studies that rely on network flows or human-curated rules, our work directly tackles the problem of objective, process-level analysis of system events. The RL agent's states are explicitly defined by nodes (processes), and actions involve selecting a causal target node ($s'$). Crucially, the rewards are meticulously structured, leveraging insights from the graph analysis regarding sparsity and centralized processes, to guide the traversal based on edge validity, process attributes (like integrity level), and network metrics, thereby creating a feasible environment for agents to learn concrete system attack path behaviors.

This framework highlights three main challenges: balancing model parameter dimensionality and efficient processing, designing a reward function that correlates sparse and low-level events with security outcomes, and strategically selecting relevant system events and user actions to model. The sparsity and navigable structure of the Cerberus Traces dataset were crucial for the agent's exploration, demonstrating the stability and scalability of our design.

In conclusion, this work demonstrates the feasibility of creating a graphical RL environment from operating system logs. It identifies effective reward design and strategic event selection as primary challenges. The \emph{SubstratumGraphEnv} serves as a vital platform for future research.

\appendix
\section{Appendices and Supplementary Material}

The following equations were used to analyze the network structures of the Cerberus Traces and BRAWL datasets. 

~\\
{\bf Edge Density ($\rho$)} for a directed graph:
\begin{equation}\rho=\frac{m}{(n(n-1))}\end{equation}
Where $m$ is the number of edges and $n$ is the number of nodes.

{\bf Clustering Coefficient: Geographic Average of Subgraph Edge Weights ($c_{u}$)}:
\begin{equation}c_{u}=\frac{1}{\deg(u)(\deg(u)-1)}\sum_{vw}\left({\hat{w}_{uv}}{\hat{w}_{uv}}{\hat{w}_{uv}}\right)^{1/3}\end{equation}

Edge weights denoted as $\hat{w}_{uv}$ are normalized by dividing the original weight $\hat{w}_{uv}$ by the maximum weight in the network $\max(w)$. That is $\hat{w}_{uv}=w_{uv}/\max(w)$. The clustering coefficient $c_u$ is defined as 0 for any node $u$ with a degree less than 2 ($\deg(u)<2$).

~\\
{\bf Average Degree Connectivity (k)} for a directed graph:
\begin{equation}k_{nn,i}^w\ \ =\frac{1}{s_i\ }\sum_{j\epsilon N\left(i\right)}{w_{ij}k_j}\end{equation}

~\\
{\bf In-degree Centrality for Nodes ($c_{in}$)}:\
\begin{equation}c_{in}(v)=\frac{{\deg}^{in}(v)}{n-1}\end{equation}

\subsection*{Author Contributions}

\begin{description}
    \item[Author 1] Conceptualization, Methodology, Analysis, Data Curation, Writing.
    \item[Author 2] Review \& Supervision.
    \item[Author 3] Review \& Supervision.
\end{description}

\nocite{*}
\bibliography{works}

\begin{thebibliography}{55}
\providecommand{\natexlab}[1]{#1}

\bibitem[{Almasan et~al.(2019)Almasan, Suárez-Varela, Rusek, Barlet-Ros, and Cabellos-Aparicio}]{almasan_deep_2019}
Almasan, P.; Suárez-Varela, J.; Rusek, K.; Barlet-Ros, P.; and Cabellos-Aparicio, A. 2019.
\newblock Deep {Reinforcement} {Learning} meets {Graph} {Neural} {Networks}: exploring a routing optimization use case.
\newblock Publisher: arXiv Version Number: 3.

\bibitem[{Applebaum et~al.(2016)Applebaum, Miller, Strom, Korban, and Wolf}]{applebaum_intelligent_2016}
Applebaum, A.; Miller, D.; Strom, B.; Korban, C.; and Wolf, R. 2016.
\newblock Intelligent, automated red team emulation.
\newblock In \emph{Proceedings of the 32nd {Annual} {Conference} on {Computer} {Security} {Applications}}, 363--373. Los Angeles California USA: ACM.
\newblock ISBN 978-1-4503-4771-6.

\bibitem[{Bland et~al.(2020)Bland, Petty, Whitaker, Maxwell, and Cantrell}]{bland_machine_2020}
Bland, J.~A.; Petty, M.~D.; Whitaker, T.~S.; Maxwell, K.~P.; and Cantrell, W.~A. 2020.
\newblock Machine {Learning} {Cyberattack} and {Defense} {Strategies}.
\newblock \emph{Computers \& Security}, 92: 101738.

\bibitem[{Botvinick et~al.(2019)Botvinick, Ritter, Wang, Kurth-Nelson, Blundell, and Hassabis}]{botvinick_reinforcement_2019}
Botvinick, M.; Ritter, S.; Wang, J.~X.; Kurth-Nelson, Z.; Blundell, C.; and Hassabis, D. 2019.
\newblock Reinforcement {Learning}, {Fast} and {Slow}.
\newblock \emph{Trends in Cognitive Sciences}, 23(5): 408--422.

\bibitem[{{C. David Hylender} et~al.(2025){C. David Hylender}, {Philippe Langlois}, {Alex Pint,}, and {Suzanne Widup}}]{c_david_hylender_2025_2025}
{C. David Hylender}; {Philippe Langlois}; {Alex Pint,}; and {Suzanne Widup}. 2025.
\newblock 2025 {Data} {Breach} {Investigations} {Report}.
\newblock Technical report, Verizon.

\bibitem[{Cheng et~al.(2025)Cheng, Wang, Li, Yin, Luan, and Shen}]{cheng_toward_2025}
Cheng, N.; Wang, X.; Li, Z.; Yin, Z.; Luan, T.~H.; and Shen, X. 2025.
\newblock Toward {Enhanced} {Reinforcement} {Learning}-{Based} {Resource} {Management} via {Digital} {Twin}: {Opportunities}, {Applications}, and {Challenges}.
\newblock \emph{IEEE Network}, 39(1): 189--196.

\bibitem[{Chowdhary et~al.(2020)Chowdhary, Huang, Mahendran, Romo, Deng, and Sabur}]{chowdhary_autonomous_2020}
Chowdhary, A.; Huang, D.; Mahendran, J.~S.; Romo, D.; Deng, Y.; and Sabur, A. 2020.
\newblock Autonomous {Security} {Analysis} and {Penetration} {Testing}.
\newblock In \emph{2020 16th {International} {Conference} on {Mobility}, {Sensing} and {Networking} ({MSN})}, 508--515. Tokyo, Japan: IEEE.
\newblock ISBN 978-1-7281-9916-0.

\bibitem[{Claypoole et~al.(2025)Claypoole, Cheung, Gehani, Yegneswaran, and Ridley}]{claypoole_interpreting_2025}
Claypoole, J.; Cheung, S.; Gehani, A.; Yegneswaran, V.; and Ridley, A. 2025.
\newblock Interpreting {Agent} {Behaviors} in {Reinforcement}-{Learning}-{Based} {Cyber}-{Battle} {Simulation} {Platforms}.
\newblock ArXiv:2506.08192 [cs].

\bibitem[{Cody et~al.(2022)Cody, Rahman, Redino, Huang, Clark, Kakkar, Kushwaha, Park, Beling, and Bowen}]{cody_discovering_2022}
Cody, T.; Rahman, A.; Redino, C.; Huang, L.; Clark, R.; Kakkar, A.; Kushwaha, D.; Park, P.; Beling, P.; and Bowen, E. 2022.
\newblock Discovering {Exfiltration} {Paths} {Using} {Reinforcement} {Learning} with {Attack} {Graphs}.
\newblock ArXiv:2201.12416 [cs].

\bibitem[{Deokar and Hazarnis(2012)}]{deokar_intrusion_2012}
Deokar, B.; and Hazarnis, A. 2012.
\newblock Intrusion {Detection} {System} using {Log} {Files} and {Reinforcement} {Learning}.
\newblock \emph{International Journal of Computer Applications}, 45(19): 28--35.

\bibitem[{Edeh, {Korki Mehdi}, and {Mekhilef, Saad}(2024)}]{edeh_reinforcement_2024}
Edeh, V.; {Korki Mehdi}; and {Mekhilef, Saad}. 2024.
\newblock Reinforcement {Learning}-{Empowered} {Graph} {Convolutional} {Network} {Framework} for {Data} {Integrity} {Attack} {Detection} in {Cyber}-{Physical} {Systems}.
\newblock \emph{CSEE Journal of Power and Energy Systems}.

\bibitem[{Elderman et~al.(2017)Elderman, J.~J.~Pater, S.~Thie, M.~Drugan, and M.~Wiering}]{elderman_adversarial_2017}
Elderman, R.; J.~J.~Pater, L.; S.~Thie, A.; M.~Drugan, M.; and M.~Wiering, M. 2017.
\newblock Adversarial {Reinforcement} {Learning} in a {Cyber} {Security} {Simulation}:.
\newblock In \emph{Proceedings of the 9th {International} {Conference} on {Agents} and {Artificial} {Intelligence}}, 559--566. Porto, Portugal: SCITEPRESS - Science and Technology Publications.
\newblock ISBN 978-989-758-219-6 978-989-758-220-2.

\bibitem[{Ergun, Sammour, and Chalhoub(2023)}]{ergun_survey_2023}
Ergun, S.; Sammour, I.; and Chalhoub, G. 2023.
\newblock A survey on how network simulators serve reinforcement learning in wireless networks.
\newblock \emph{Computer Networks}, 234: 109934.

\bibitem[{{Farama Foundation}(2025{\natexlab{a}})}]{farama_foundation_composite_nodate}
{Farama Foundation}. 2025{\natexlab{a}}.
\newblock Composite {Spaces}.

\bibitem[{{Farama Foundation}(2025{\natexlab{b}})}]{farama_foundation_fundamental_nodate}
{Farama Foundation}. 2025{\natexlab{b}}.
\newblock Fundamental {Spaces}.

\bibitem[{{Farama Foundation}(2025{\natexlab{c}})}]{farama_foundation_source_nodate}
{Farama Foundation}. 2025{\natexlab{c}}.
\newblock Source code for gymnasium.spaces.graph.

\bibitem[{{Fritz Obermeyer}(2017)}]{fritz_obermeyer_pytorchtorchdistributionscategoricalpy_2017}
{Fritz Obermeyer}. 2017.
\newblock pytorch/torch/distributions/categorical.py at v2.7.0 · pytorch/pytorch.

\bibitem[{Gueriani, Kheddar, and Mazari(2024)}]{gueriani_deep_2024}
Gueriani, A.; Kheddar, H.; and Mazari, A.~C. 2024.
\newblock Deep {Reinforcement} {Learning} for {Intrusion} {Detection} in {IoT}: {A} {Survey}.
\newblock Publisher: arXiv Version Number: 1.

\bibitem[{Harries et~al.(2020)Harries, Clarke, Chapman, Nallamalli, Ozgur, Jain, Leung, Lim, Dietrich, Hernández-Lobato, Ellis, Zhang, and Ciosek}]{harries_drift_2020}
Harries, L.; Clarke, R.~S.; Chapman, T.; Nallamalli, S. V. P. L.~N.; Ozgur, L.; Jain, S.; Leung, A.; Lim, S.; Dietrich, A.; Hernández-Lobato, J.~M.; Ellis, T.; Zhang, C.; and Ciosek, K. 2020.
\newblock {DRIFT}: {Deep} {Reinforcement} {Learning} for {Functional} {Software} {Testing}.
\newblock ArXiv:2007.08220 [cs].

\bibitem[{Ispahany et~al.(2025)Ispahany, Islam, Khan, and Islam}]{ispahany_sysmon_2025}
Ispahany, J.; Islam, M.~R.; Khan, M.~A.; and Islam, M.~Z. 2025.
\newblock A {Sysmon} {Incremental} {Learning} {System} for {Ransomware} {Analysis} and {Detection}.
\newblock ArXiv:2501.01089 [cs].

\bibitem[{Jahin et~al.(2025)Jahin, Soudeep, Farid, Mridha, Kabir, Islam, and Karim}]{jahin_cagn-gat_2025}
Jahin, M.~A.; Soudeep, S.; Farid, F.~A.; Mridha, M.~F.; Kabir, R.; Islam, M.~R.; and Karim, H.~A. 2025.
\newblock {CAGN}-{GAT} {Fusion}: {A} {Hybrid} {Contrastive} {Attentive} {Graph} {Neural} {Network} for {Network} {Intrusion} {Detection}.
\newblock ArXiv:2503.00961 [cs].

\bibitem[{Jamshidi et~al.(2025)Jamshidi, Nikanjam, Nafi, Khomh, and Rasta}]{jamshidi_application_2025}
Jamshidi, S.; Nikanjam, A.; Nafi, K.~W.; Khomh, F.; and Rasta, R. 2025.
\newblock Application of {Deep} {Reinforcement} {Learning} for {Intrusion} {Detection} in {Internet} of {Things}: {A} {Systematic} {Review}.
\newblock \emph{Internet of Things}, 31: 101531.
\newblock ArXiv:2504.14436 [cs].

\bibitem[{{Jeff Hancock}(2022)}]{jeff_hancock_psychology_2022}
{Jeff Hancock}. 2022.
\newblock The {Psychology} of {Human} {Error}: {Understanding} the mistakes that compromise your company's cybersecurity.
\newblock Technical report, Tessian Limited, Stanford University.

\bibitem[{Kenyon, Deka, and Elizondo(2020)}]{kenyon_are_2020}
Kenyon, A.; Deka, L.; and Elizondo, D. 2020.
\newblock Are public intrusion datasets fit for purpose characterising the state of the art in intrusion event datasets.
\newblock \emph{Computers \& Security}, 99: 102022.

\bibitem[{Kujanpää, Victor, and Ilin(2021)}]{kujanpaa_automating_2021}
Kujanpää, K.; Victor, W.; and Ilin, A. 2021.
\newblock Automating {Privilege} {Escalation} with {Deep} {Reinforcement} {Learning}.
\newblock In \emph{Proceedings of the 14th {ACM} {Workshop} on {Artificial} {Intelligence} and {Security}}, 157--168. Virtual Event Republic of Korea: ACM.
\newblock ISBN 978-1-4503-8657-9.

\bibitem[{Kulkarni and OShaughnessy(2025)}]{kulkarni_malware_2025}
Kulkarni, P.; and OShaughnessy, S. 2025.
\newblock Malware {Detection} {Using} {Dynamic} {Graph} {Neural} {Networks}.
\newblock \emph{European Conference on Cyber Warfare and Security}, 24(1): 830--837.

\bibitem[{Li, Shi, and Leeuwen(2024)}]{li_graph_2024}
Li, Z.; Shi, J.; and Leeuwen, M.~v. 2024.
\newblock Graph {Neural} {Networks} based {Log} {Anomaly} {Detection} and {Explanation}.
\newblock ArXiv:2307.00527 [cs].

\bibitem[{Liu et~al.(2025)Liu, Su, Li, Huang, and Li}]{liu_digital_2025}
Liu, H.; Su, W.; Li, T.; Huang, W.; and Li, Y. 2025.
\newblock Digital {Twin} {Enhanced} {Multi}-{Agent} {Reinforcement} {Learning} for {Large}-{Scale} {Mobile} {Network} {Coverage} {Optimization}.
\newblock \emph{ACM Transactions on Knowledge Discovery from Data}, 19(1): 1--23.

\bibitem[{{Los Alamos National Laboratory}(2024{\natexlab{a}})}]{los_alamos_national_laboratory_average_degree_connectivity_nodate}
{Los Alamos National Laboratory}. 2024{\natexlab{a}}.
\newblock average\_degree\_connectivity — {NetworkX} 3.4.2 documentation.

\bibitem[{{Los Alamos National Laboratory}(2024{\natexlab{b}})}]{los_alamos_national_laboratory_clustering_nodate}
{Los Alamos National Laboratory}. 2024{\natexlab{b}}.
\newblock clustering — {NetworkX} 3.4.2 documentation.

\bibitem[{{Los Alamos National Laboratory}(2024{\natexlab{c}})}]{los_alamos_national_laboratory_density_nodate}
{Los Alamos National Laboratory}. 2024{\natexlab{c}}.
\newblock density — {NetworkX} 3.4.2 documentation.

\bibitem[{{Los Alamos National Laboratory}(2024{\natexlab{d}})}]{los_alamos_national_laboratory_in_degree_centrality_nodate}
{Los Alamos National Laboratory}. 2024{\natexlab{d}}.
\newblock in\_degree\_centrality — {NetworkX} 3.4.2 documentation.

\bibitem[{López-Montero et~al.(2025)López-Montero, Álvarez Aldana, Morales-Martínez, Gil-López, and García}]{lopez-montero_reinforcement_2025}
López-Montero, D.; Álvarez Aldana, J.~L.; Morales-Martínez, A.; Gil-López, M.; and García, J. M.~A. 2025.
\newblock Reinforcement {Learning} for {Automated} {Cybersecurity} {Penetration} {Testing}.
\newblock Version Number: 1.

\bibitem[{Maeda and Mimura(2021)}]{maeda_automating_2021}
Maeda, R.; and Mimura, M. 2021.
\newblock Automating post-exploitation with deep reinforcement learning.
\newblock \emph{Computers \& Security}, 100: 102108.

\bibitem[{Marbel et~al.(2024)Marbel, Cohen, Dubin, Dvir, and Hajaj}]{marbel_cloudy_2024}
Marbel, R.; Cohen, Y.; Dubin, R.; Dvir, A.; and Hajaj, C. 2024.
\newblock Cloudy with a {Chance} of {Anomalies}: {Dynamic} {Graph} {Neural} {Network} for {Early} {Detection} of {Cloud} {Services}' {User} {Anomalies}.
\newblock ArXiv:2409.12726 [cs].

\bibitem[{{Mike Kemmerer} and {Craig Wampler}(2018)}]{mike_kemmerer_mitrebrawl-public-game-001_2018}
{Mike Kemmerer}; and {Craig Wampler}. 2018.
\newblock mitre/brawl-public-game-001.
\newblock Original-date: 2017-08-29T15:03:50Z.

\bibitem[{Mitra et~al.(2024)Mitra, Chakraborty, Neupane, Piplai, and Mittal}]{mitra_use_2024}
Mitra, S.; Chakraborty, T.; Neupane, S.; Piplai, A.; and Mittal, S. 2024.
\newblock Use of {Graph} {Neural} {Networks} in {Aiding} {Defensive} {Cyber} {Operations}.
\newblock ArXiv:2401.05680 [cs].

\bibitem[{Mnih et~al.(2016)Mnih, Badia, Mirza, Graves, Lillicrap, Harley, Silver, and Kavukcuoglu}]{mnih_asynchronous_2016}
Mnih, V.; Badia, A.~P.; Mirza, M.; Graves, A.; Lillicrap, T.~P.; Harley, T.; Silver, D.; and Kavukcuoglu, K. 2016.
\newblock Asynchronous {Methods} for {Deep} {Reinforcement} {Learning}.
\newblock ArXiv:1602.01783 [cs].

\bibitem[{Molina-Markham et~al.(2021)Molina-Markham, Miniter, Powell, and Ridley}]{molina-markham_network_2021}
Molina-Markham, A.; Miniter, C.; Powell, B.; and Ridley, A. 2021.
\newblock Network {Environment} {Design} for {Autonomous} {Cyberdefense}.
\newblock ArXiv:2103.07583 [cs].

\bibitem[{Munikoti et~al.(2022)Munikoti, Agarwal, Das, Halappanavar, and Natarajan}]{munikoti_challenges_2022}
Munikoti, S.; Agarwal, D.; Das, L.; Halappanavar, M.; and Natarajan, B. 2022.
\newblock Challenges and {Opportunities} in {Deep} {Reinforcement} {Learning} with {Graph} {Neural} {Networks}: {A} {Comprehensive} review of {Algorithms} and {Applications}.
\newblock ArXiv:2206.07922 [cs].

\bibitem[{Pratomo(2023)}]{pratomo_bazz-066cerberus-trace_2023}
Pratomo, B.~A. 2023.
\newblock bazz-066/cerberus-trace.
\newblock Original-date: 2023-07-06T11:17:46Z.

\bibitem[{{Proofpoint, Inc.}(2024)}]{proofpoint_inc_proofpoints_2024}
{Proofpoint, Inc.} 2024.
\newblock Proofpoint’s 2024 {Voice} of the {CISO} {Report} {Reveals} that {Three}-{Quarters} of {CISOs} {Identify} {Human} {Error} as {Leading} {Cybersecurity} {Risk}.
\newblock Technical report.

\bibitem[{{PyTorch Foundation}(2025)}]{pytorch_foundation_linear_nodate}
{PyTorch Foundation}. 2025.
\newblock Linear — {PyTorch} 2.7 documentation.

\bibitem[{{Richard Barnes}(2021)}]{richard_barnes_torchnnfunctionalmse_loss_2021}
{Richard Barnes}. 2021.
\newblock torch.nn.functional.mse\_loss — {PyTorch} 2.7 documentation.

\bibitem[{Sahu, Venkatraman, and Macwan(2023)}]{sahu_reinforcement_2023}
Sahu, A.; Venkatraman, V.; and Macwan, R. 2023.
\newblock Reinforcement {Learning} {Environment} for {Cyber}-{Resilient} {Power} {Distribution} {System}.
\newblock \emph{IEEE Access}, 11: 127216--127228.

\bibitem[{{SentinelOne}(2025)}]{sentinelone_50_2025}
{SentinelOne}. 2025.
\newblock 50+ {Cloud} {Security} {Statistics} in 2025.
\newblock Technical report.

\bibitem[{Terranova, {Abdelkader Lahmadi}, and Chrisment(2024)}]{terranova_deep_2024}
Terranova, F.; {Abdelkader Lahmadi}; and Chrisment, I. 2024.
\newblock Deep {Reinforcement} {Learning} for {Automated} {Cyber}-{Attack} {Path} {Prediction} in {Communication} {Networks}.
\newblock Publisher: Unpublished.

\bibitem[{Toyama et~al.(2021)Toyama, Hamel, Gergely, Comanici, Glaese, Ahmed, Jackson, Mourad, and Precup}]{toyama_androidenv_2021}
Toyama, D.; Hamel, P.; Gergely, A.; Comanici, G.; Glaese, A.; Ahmed, Z.; Jackson, T.; Mourad, S.; and Precup, D. 2021.
\newblock {AndroidEnv}: {A} {Reinforcement} {Learning} {Platform} for {Android}.
\newblock ArXiv:2105.13231 [cs].

\bibitem[{{Vincent Moens}(2024)}]{vincent_moens_bug_nodate}
{Vincent Moens}. 2024.
\newblock [{BUG}] {Incompatibility} between {TorchRL} and {Gymnasium} 1.0: {Auto}-{Reset} {Feature} {Breaks} {Modularity} and {Data} {Integrity}.

\bibitem[{Wang et~al.(2025)Wang, Zheng, Gui, Hua, and Hassan}]{wang_are_2025}
Wang, C.; Zheng, P.; Gui, J.; Hua, C.; and Hassan, W.~U. 2025.
\newblock Are {We} {There} {Yet}? {Unraveling} the {State}-of-the-{Art} {Graph} {Network} {Intrusion} {Detection} {Systems}.
\newblock ArXiv:2503.20281 [cs].

\bibitem[{Xie, Zhang, and Babar(2022)}]{xie_loggddetecting_2022}
Xie, Y.; Zhang, H.; and Babar, M.~A. 2022.
\newblock {LogGD}:{Detecting} {Anomalies} from {System} {Logs} by {Graph} {Neural} {Networks}.
\newblock ArXiv:2209.07869 [cs].

\bibitem[{Xu and Xie(2005)}]{hutchison_reinforcement_2005}
Xu, X.; and Xie, T. 2005.
\newblock A {Reinforcement} {Learning} {Approach} for {Host}-{Based} {Intrusion} {Detection} {Using} {Sequences} of {System} {Calls}.
\newblock In Hutchison, D.; Kanade, T.; Kittler, J.; Kleinberg, J.~M.; Mattern, F.; Mitchell, J.~C.; Naor, M.; Nierstrasz, O.; Pandu~Rangan, C.; Steffen, B.; Sudan, M.; Terzopoulos, D.; Tygar, D.; Vardi, M.~Y.; Weikum, G.; Huang, D.-S.; Zhang, X.-P.; and Huang, G.-B., eds., \emph{Advances in {Intelligent} {Computing}}, volume 3644, 995--1003. Berlin, Heidelberg: Springer Berlin Heidelberg.
\newblock ISBN 978-3-540-28226-6 978-3-540-31902-3.
\newblock Series Title: Lecture Notes in Computer Science.

\bibitem[{Yang and Liu(2022)}]{yang_behaviour-diverse_2022}
Yang, Y.; and Liu, X. 2022.
\newblock Behaviour-{Diverse} {Automatic} {Penetration} {Testing}: {A} {Curiosity}-{Driven} {Multi}-{Objective} {Deep} {Reinforcement} {Learning} {Approach}.
\newblock ArXiv:2202.10630 [cs].

\bibitem[{Zhao, Queralta, and Westerlund(2020)}]{zhao_sim--real_2020}
Zhao, W.; Queralta, J.~P.; and Westerlund, T. 2020.
\newblock Sim-to-{Real} {Transfer} in {Deep} {Reinforcement} {Learning} for {Robotics}: a {Survey}.
\newblock In \emph{2020 {IEEE} {Symposium} {Series} on {Computational} {Intelligence} ({SSCI})}, 737--744. Canberra, ACT, Australia: IEEE.
\newblock ISBN 978-1-7281-2547-3.

\bibitem[{Zheng, Wang, and Song(2021)}]{zheng_opengraphgym-mg_2021}
Zheng, W.; Wang, D.; and Song, F. 2021.
\newblock {OpenGraphGym}-{MG}: {Using} {Reinforcement} {Learning} to {Solve} {Large} {Graph} {Optimization} {Problems} on {MultiGPU} {Systems}.
\newblock ArXiv:2105.08764 [cs].

\end{thebibliography}

\end{document}